\documentclass[12pt]{article}

\usepackage{amssymb}
\usepackage{amsmath}
\usepackage{amsfonts}
\usepackage{graphicx}
\usepackage{mathrsfs}
\usepackage{comment}
\usepackage{cite}
\usepackage{color}
\begin{document}
\def\ebq{\end{equation} \begin{equation}}
\renewcommand{\figurename}{Fig.}
\renewcommand{\tablename}{Table.}
\newcommand{\Slash}[1]{{\ooalign{\hfil#1\hfil\crcr\raise.167ex\hbox{/}}}}
\newcommand{\bra}[1]{ \langle {#1} | }
\newcommand{\ket}[1]{ | {#1} \rangle }
\newcommand{\beq}{\begin{equation}}  \newcommand{\eeq}{\end{equation}}
\newcommand{\bef}{\begin{figure}}  \newcommand{\eef}{\end{figure}}
\newcommand{\bec}{\begin{center}}  \newcommand{\eec}{\end{center}}
\newcommand{\non}{\nonumber}  \newcommand{\eqn}[1]{\begin{equation} {#1}\end{equation}}
\newcommand{\laq}[1]{\label{eq:#1}}  
\newcommand{\dd}[1]{{d \o d{#1}}}
\newcommand{\Eq}[1]{Eq.~(\ref{eq:#1})}
\newcommand{\Eqs}[1]{Eqs.~(\ref{eq:#1})}
\newcommand{\eq}[1]{(\ref{eq:#1})}
\newcommand{\Sec}[1]{Sec.\ref{chap:#1}}
\newcommand{\ab}[1]{\left|{#1}\right|}
\newcommand{\vev}[1]{ \left\langle {#1} \right\rangle }
\newcommand{\bs}[1]{ {\boldsymbol {#1}} }
\newcommand{\lac}[1]{\label{chap:#1}}
\newcommand{\SU}[1]{{\rm SU{#1} } }
\newcommand{\SO}[1]{{\rm SO{#1}} }

\def\({\left(}
\def\){\right)}
\def\dt{{d \o dt}}
\def\diag{\mathop{\rm diag}\nolimits}
\def\Spin{\mathop{\rm Spin}}
\def\O{\mathcal{O}}
\def\U{\mathop{\rm U}}
\def\Sp{\mathop{\rm Sp}}
\def\SL{\mathop{\rm SL}}
\def\tr{\mathop{\rm tr}}
\newcommand{\OR}{~{\rm or}~}
\newcommand{\AND}{~{\rm and}~}
\newcommand{\EV}{ {\rm \, eV} }
\newcommand{\KEV}{ {\rm \, keV} }
\newcommand{\MEV}{ {\rm \, MeV} }
\newcommand{\GEV}{ {\rm \, GeV} }
\newcommand{\TEV}{ {\rm \, TeV} }

\def\o{\over}
\def\a{\alpha}
\def\b{\beta}
\def\c{\varepsilon}
\def\d{\delta}
\def\e{\epsilon}
\def\f{\phi}
\def\g{\gamma}
\def\h{\theta}
\def\k{\kappa}
\def\l{\lambda}
\def\m{\mu}
\def\n{\nu}
\def\p{\psi}
\def\q{\partial}
\def\r{\rho}
\def\s{\sigma}
\def\t{\tau}
\def\u{\upsilon}
\def\v{\varphi}
\def\w{\omega}
\def\x{\xi}
\def\y{\eta}
\def\z{\zeta}
\def\D{\Delta}
\def\G{\Gamma}
\def\H{\Theta}
\def\L{\Lambda}
\def\F{\Phi}
\def\P{\Psi}
\def\S{\Sigma}
\def\me{\mathrm e}
\def\ol{\overline}
\def\tl{\tilde}
\def\*{\dagger}

\begin{center}

\hfill KEK--TH--2153\\

\vspace{1.5cm}

{\Large\bf Throwing away antimatter via neutrino oscillations during the reheating era}

\vspace{1.5cm}

{\bf Shintaro Eijima$^{\rm (a)}$, Ryuichiro Kitano$^{\rm (a,b)}$ and Wen Yin$^{\rm (c)}$}

{
\vskip 0.1in
$^{\rm (a)}${\it
Theory Center, IPNS, KEK, Tsukuba, Ibaraki 305-0801, Japan}

\vskip 0.1in
$^{\rm (b)}${\it
Graduate University for Advanced Studies (Sokendai), Tsukuba 305-0801, Japan}

\vskip 0.1in

$^{\rm (c)}${\it Department of Physics, KAIST, Daejeon 34141, Korea}

}
\vspace{12pt}
\vspace{1.5cm}

\date{\today $\vphantom{\bigg|_{\bigg|}^|}$}

\abstract{
The simplest possibility to explain the baryon asymmetry of the Universe
is to assume that radiation is created asymmetrically between baryons
and anti-baryons after the inflation.
We propose a new mechanism of this kind where 
CP-violating flavor oscillations of left-handed leptons in the reheating era distribute the lepton asymmetries
partially into the right-handed neutrinos while net asymmetry is not
created.
The asymmetry stored in the right-handed neutrinos is later washed out
by the lepton number violating decays, and it ends up with the net
lepton asymmetry in the Standard Model particles, which is converted
into the baryon asymmetry by the sphaleron process.
This scenario works for a range of masses of the right-handed neutrinos
while no fine-tuning among the masses is required.
The reheating temperature of the Universe can be as low as $\O(10)$~TeV
if we assume that the decays of inflatons in the perturbative regime are
responsible for the reheating. For the case of the reheating via the
dissipation effects, the reheating temperature can be as low as
$\O(100)$~GeV.
 }
\end{center}
\clearpage

\setcounter{page}{1}
\setcounter{footnote}{0}

\setcounter{footnote}{0}
\section{Introduction}
Missing antimatter is one of the mysteries in the history of the Universe.
The baryon asymmetry cannot be the initial condition in the inflationary
cosmology while the thermal history within the Standard Model of
particle physics seems to fail to explain it. It is plausible that the
neutrino masses may be something to do with this mystery since the
Majorana masses of neutrinos together with the sphaleron process
provides us with a new source of the baryon number violation as well as
CP violation~\cite{Fukugita:1986hr}.

It is well-known that three conditions need to be satisfied for the
creation of the baryon asymmetry after inflation~\cite{Sakharov:1967dj}:
baryon number violation, CP violation, and a stage with
out-of-equilibrium.
Recently it has been shown that all the three conditions can be
satisfied at the very beginning of the Universe in the Standard Model
with a dimension five operator to generate the neutrino Majorana
masses~\cite{Hamada:2016oft, Hamada:2018epb}.  (See also the reheating
era baryogenesis~\cite{Hamada:2015xva, Hamada:2016npz, Hamada:2016oft}, and non-thermal leptogenesis~\cite{Lazarides:1991wu, Asaka:1999yd, Hamaguchi:2001gw}.)
In these scenarios, the right-handed neutrinos are not necessary, and
thus the effective theory to describe the phenomena is the same as the
one for the low energy experiments, such as the neutrino oscillation
experiments as well as the neutrinoless double beta decays. This
provides us with tight connections between the baryon asymmetry of the
Universe and the low energy experiments.

The key fact is that flavor oscillations of active neutrinos during the
reheating era provide CP violation just as in the neutrino oscillation
phenomena we observe today~\cite{Hamada:2016oft}. It was shown that only
with adding the higher dimensional Majorana mass term, $LLHH$, to the
renormalizable SM, the baryon asymmetry can be generated during the
thermalization process~\cite{Hamada:2018epb}.  The oscillations are
induced due to the misalignment of the eigenbasis of the effective mass
matrices, governed by the matter effects, and that of the $LLHH$ interactions.
The observed baryon asymmetry is shown to be explained if the reheating
temperature is higher than $10^8\GEV$.  
The scenario works for whatever mechanism for the generation of neutrino
Majorana masses at sufficiently high renormalization scale.

The baryogenesis with two or three generations of the right-handed
neutrinos have been studied widely in connection with 
the generation of the neutrino masses by the seesaw mechanism~\cite{Yanagida:1979as, 
GellMann:1980vs, Yanagida:1980xy, Schechter:1980gr,
Schechter:1981cv}. (See also Refs.~\cite{Minkowski:1977sc, Mohapatra:1979ia})
Thermal leptogenesis~\cite{Fukugita:1986hr} assumes the thermal bath
of the SM including the right-handed neutrinos as the initial condition,
and the lepton asymmetry is produced by the out-of-equilibrium decays of
right-handed neutrinos, which requires $T_R\gtrsim 10^8\GEV$. (See
Refs.~\cite{Buchmuller:2005eh, Davidson:2008bu} for reviews.)  It has
been shown that the reheating temperature can be lower if one tunes the
difference between the right-handed neutrino masses so that the resonant
effects take place~\cite{Pilaftsis:1997dr,Buchmuller:1997yu}.
The right-handed neutrinos can be much lighter than $\O(100)\GEV$ when
the flavor oscillations of the right-handed neutrinos get important
while some tuning in the mass spectrum is
necessary~\cite{Akhmedov:1998qx,Asaka:2005pn}.
In this scenario, the abundance of the right-handed neutrinos, which is
assumed to be zero at the beginning of the Universe, are generated through
the scattering of the left-handed leptons.  
The lepton asymmetries originated from the oscillation among
right-handed neutrinos are stored separately into the left-handed and
right-handed neutrino sectors.
In the case of the neutrinos with the Dirac masses, one can also
consider the possibility that the lepton asymmetry is stored in the
right-handed neutrinos~\cite{Dick:1999je}.

In this paper, we consider flavor oscillations of active neutrinos
during the reheating era in the seesaw model~\cite{Yanagida:1979as, 
GellMann:1980vs, Yanagida:1980xy, Schechter:1980gr,
Schechter:1981cv}.
Through the Yukawa interactions between the lepton
doublets and the right-handed neutrinos, the CP-violating flavor
oscillations distribute lepton asymmetries into the left-handed and
right-handed neutrinos, while total lepton asymmetry is conserved.
The lepton asymmetry stored in the left-handed leptons is, in turn,
converted into the baryon asymmetry by the sphaleron process.
If the right-handed neutrinos never come into the thermal equilibrium
until the sphaleron process shuts off at $T \sim 100\GEV$, the created
baryon asymmetry remains today.
In the scenario where the reheating is caused by the perturbative decay
of the inflaton, the reheating temperature should satisfy $T_R\gtrsim
7\TEV$ for the successful baryogenesis.  We also discuss the
possibility that the reheating is due to the dissipation effect. In that
case the right-handed neutrino can be lighter than $100\GEV$ and the
reheating temperature can be as low as $T_R \sim 100 \GEV$.

The new mechanism does not require a fine-tuning of the mass
degeneracy. 
Since the density matrices of the initial left-handed neutrinos are not
in the thermal ones in the reheating era, the asymmetry via oscillation
is produced at the leading order in the perturbation of the neutrino
Yukawa couplings.  As a result, large enough baryon asymmetry can be
produced. The mechanism works with a
single right-handed neutrino, and thus no tuning among masses of
right-handed neutrinos is necessary.

This paper is organized as follows. 
The main idea is shown in the \Sec{Sec2} by assuming the inflaton decays in the perturbative regime. 
In \Sec{Sec3}, we discuss the case with reheating via the dissipation processes. 
The last section is devoted to conclusions and discussion.

\section{Active neutrino oscillation for baryogenesis}

\lac{Sec2}
We introduce a singlet fermion to the Standard Model gauge group, $N$, which is one
of three right-handed neutrinos
responsible for the seesaw mechanism. For a while, we ignore its mass.
The Lagrangian is given as
\beq
\laq{1}
{\cal L}\supset -y_{N i} \tilde{H}^*  \bar{N} \hat{P}_L L_i, 
\eeq
where $y_{N i}$ are the Yukawa coupling constants, $L_i$ $(i=e,\m,\t)$
is the lepton doublet field, $\hat{P}_L$ is a left-handed projection operator and $\tilde{H} \equiv i \sigma_2 H$ is the
Higgs field.
We restrict ourselves in the case for $|y_{N_i}| \ll \O(1)$.
Here we take the basis that $y_{N i}$ is real by making the phase
rotation of $L_e, L_\m$ and $N$ without loss of generality.

\subsection{Inflaton decay in the perturbative regime}
We introduce an inflaton field, $\f$, which once dominates over the
Universe.
The mass is $m_{\f}.$ Let us first assume that the reheating of the
Universe proceeds via the $\f$ perturbative decays for simplicity.  We
suppose that the decay has some branching fraction $B$ to the active
neutrinos, 
\beq 
\f\rightarrow L_{ \f}+X, \quad
\bar{L}_{\f}+\bar{X}. \laq{proc}
\eeq 
Here $X$ denotes arbitrary final states, and $B\leq 1$. The final lepton
state, $L_\phi$, is in general a linear combination of $L_e$, $L_\mu$,
and $L_\tau$.
Through the dominant decay channels of $\f$ the Universe is reheated to
the temperature, $T=T_R\simeq ({g_* \pi^2 / 90 })^{-1/4}\sqrt{\G M_P},$
with the total decay width $\G$, the effective relative degrees of
freedom $g_*\simeq 106.75$, and the reduced Planck mass $M_P\simeq 2.4\times
10^{18}\GEV$.

At the moment of the inflaton decay $t=t_R\equiv 1/\G$, there are two
components in the Universe. One is the thermal plasma which
is generated at the preheating era at $t<t_R.$\footnote{The produced baryon asymmetry via the thermalization at $t<t_R$ is suppressed due to the dilution.} The thermal distribution
is characterized by the temperature $T_R$, which should satisfy
\beq
T_R\leq m_\phi
\eeq
for the regime of the perturbative decay.  Another component is the
direct decay product at $t=t_R$ which includes the active leptons,
$L_\f$. These leptons are generally out of equilibrium. For instance, if
we consider a two-body decay to the lepton, the component includes
monochromatic modes of the leptons with energy around $m_\f/2.$ The
lepton will be thermalized promptly due to the interaction with the
thermal plasma.

In the following, we will discuss the leptogenesis via the active
lepton/neutrino oscillations during this rapid thermalization process
and show that this scenario works with low reheating temperature if
there is a sufficient amount of CP violation.
The lepton asymmetries are divided into two sectors: the active neutrino
sector, $\D_{\rm vis} $, and $N$ sector, $\D_{N}$, while the total
asymmetry is zero, i.e. $\D_{\rm vis}+\D_{N}=0$, due to the conservation
of the lepton number once we ignore the mass term of $N$.
After the thermalization of left-handed leptons only $\D_{\rm vis}$ is
important and can be converted into the baryon asymmetry by the
sphaleron process.  If $N$ is not thermalized
until the temperature drops to $T< T_{\rm sph}\sim 100\GEV$, where the sphaleron process
freezes out, $\D_N$ would not be transferred back into the visible
sector. As a result, the produced baryon asymmetry is maintained until
today.

The asymmetry $\Delta_N$ is produced in the following way.  At $t=t_R$,
the lepton of momentum $\bf p$ produced by an inflaton decay, $L_{\f}$,
is represented as a quantum state
\beq
\ket{L_{\f}, t_R},
\eeq
which evolves as 
\beq
\ket{ L_{\f}, t}= \sum_i c_i \exp \left[ -i\int^{t}_{t_R}{ E_i d t^{\prime}} \right] \ket{i} 
\label{eq:Eint}
\eeq
where $\ket{i}$ is the flavor eigenstate of the left-handed leptons, $i=e,\m,\t$, with momentum $p$ which is around $m_\f$.   We have defined 
\beq
\laq{ph}
c_i\equiv \langle{i}\ket{L_{\f},t_R}.
\eeq
Here $c_\t$ can be taken to be real by the field redefinition of $L_\t$
without loss of generality, but $c_e$ and $c_\m$ are in general complex
numbers. The flavor oscillation phenomena happen through thermal
potentials, which are created by the preexisting thermal plasma. For
$|{\bf p}| \gtrsim T$, the dispersion relation becomes flavor dependent such
as
\beq
E_i\simeq y_i^2 {T^2\over 16|{\bf p}|}+ \cdots , \quad (i = e, \mu, \tau),
\eeq
where $y_i$ are the Yukawa coupling constant for the charged leptons,
$y_i = m_i / \langle H \rangle$.
We assumed $y_\t \gg y_N$, and `$\cdots$' contains the the flavor-blind
terms irrelevant for the flavor oscillation.

The thermal plasma plays two important roles. One is to induce the
thermal potential for the flavor oscillation as just discussed.  The other
is that it prevents the flavor oscillation from lasting too long. The
oscillation is terminated when the leptons annihilate with the plasma.
The free propagation time scale, $t_{\rm MFP},$ is given approximately as the inverse of the thermalization rate,
\beq
\laq{MFP}
t_{\rm MFP}\simeq \G_{\rm th}^{-1} \simeq \(\a_2^2 T\sqrt{\frac{T}{|{\bf p}|}}\)^{-1}.
\eeq
where we have taken into account the Landau-Pomeranchuk-Migdal (LPM)
effects~\cite{Landau:1953um,Migdal:1956tc} for estimating  the energy loss process important for the thermalization. 
The inelastic scattering rate via a $t$-channel gauge boson exchange is na\"{i}vely $\O(\a_2^2 T)$.
However, at the quantum level, one must take into account the coherent multiple gauge boson emissions, when an energetic lepton is injected into the medium. This effect leads to the suppression factor $\sqrt{T/|{\bf p}|}.$

The leptons from the inflaton decays lose the energy and settle down to
a state with $|{\bf p}| \sim T$ after traveling in the plasma for a
typical time scale $t_{\rm MFP}$.
The scattering via gauge interactions does not touch the flavor and the
flavor oscillation continues even after the scattering.  It is the pair
annihilation of the leptons via gauge interactions that terminates the
oscillation. It happens most effectively after the energy of the lepton
drops down to $|{\bf p}|\sim T= T_R.$ The time scale of the pair
annihilation is given as
\beq
(\D t_{\rm pair})^{-1}\sim \G_{\rm pair}\sim \a_2^2 T,
\eeq
which is even shorter than the time scale of the thermalization, $t_{\rm
MFP}$.
The flavor oscillation is also the most effective for $|{\bf p}| \sim
T$. Therefore, the quantum state of the leptons shortly after the time
scale, $t_{\rm MFP}$, is given by
\beq
\laq{ini}
\ket{ L_{\f},t_R+t_{\rm MFP}}\simeq \sum_i c_i \exp{\left[-i \frac{y_i^2}{16\a_2^2}+...\right]}\ket{i}.
\eeq
The integration in Eq.~\eqref{eq:Eint} is approximated by $E_i \Delta
t_{\rm pair}$ evaluated at $|{\bf p}| = T$.
The evolution of each flavor component differs by a phase, and for
$\tau$ the difference is,
\beq
\frac{y_\tau^2}{16 \a_2^2}\sim 0.005.
\eeq
This can be the origin of the baryon asymmetry $\O(10^{-10})$. The
effects are not suppressed by a ratio of the neutrino masses or charged
lepton masses to the energy scale of the problem, $m_\phi$ or $T_R$. The
matter effects in the finite temperature plasma make it possible to
induce the large quantum oscillation phenomenon.
We emphasize here that even though the oscillation is stopped by the
time scale of the pair annihilations, the density matrices in the flavor
space are still not collapsed into the flavor eigenbasis until the
Yukawa interactions get important.

After the evolution of the quantum state, the flavor is ``observed'' by
the flavor dependent interaction with the thermal plasma.  
At this stage, the lepton state is identified as one of the flavors,
$e$, $\mu$ or $\tau$ by the Yukawa interactions of the charged leptons.
As a rare process, however, the flavor can be ``observed'' by the
neutrino Yukawa interaction in Eq.~\eq{1}.
The observation through the Yukawa interaction of Eq.~\eq{1} happens at
the probability of
\beq
\eta \sim \frac{\sum{|c_i|^2 \sigma_{  \n_i t_L\to N t_R}}}{\sum{\ab{c_i}^2 \s_{\n_i t_L \to \t_R b_R }}}\sim \frac{|y_{N}|^2}{y_\t^2},
\eeq
where we have defined $|y_N|^2\equiv \sum_{i} |y_{N_i}|^2$. 
The probability is normalized by the process with the largest cross
section, i.e., the scattering via $y_\tau$.

As in the ordinary neutrino oscillation this rare process can have CP
asymmetry since there are strong phases (CP-even phases) from the
oscillation and the CP-odd phases in the new interactions including the
inflaton couplings. This is because we cannot remove all of the CP
phases from the field redefinition as we have performed. The CP asymmetry in
the probability is given by
\beq
P_{L_\f\to L_N}-P_{\overline{L}_\f \to \overline{L}_N}= \eta\(|\bra{  L_N} L_{\f}, t_R+t_{\rm MFP}  \rangle |^2-|\bra{ \ol{ L_N}}\ol{L}_{\f}, t_R+t_{\rm MFP}  \rangle |^2\).
\eeq
Here we have defined the state $\bra{L_N}$ as the eigenstate in the interaction basis of Eq.~\eq{1} which satisfies $\bra{L_N}i\rangle=y_{Ni}/|y_N|.$
Thus, 
\beq
\bra{ L_N} {L_{\f},t_R+ t_{\rm MFP}}\rangle \simeq \sum_{i}c_i \exp{\left[i \frac{y_{i}^2}{16\a_2^2}+...\right] }{y_{Ni}\over |y_N|}.
\eeq 
The probability is estimated as 
\beq
P_{L_\f\to L_N}-P_{\overline{L}_\f \to \overline{L}_N} \simeq \sum_{i>j}4\Im{[c_i c_j^*]} \sin{\(\frac{y_i^2-y_j^2}{16\a_2^2}\)}\frac{y_{N_i} y_{N_j}}{y_\t^2}\sim {c_\t  {y_{N_\t} \sum_{i=e,\m}{\Im[c^*_i y_{N_i}]}}\over 4\a_2^2}.
\eeq
The leptonic asymmetry in $N$ is produced with this probability for each
leptons generated by the inflaton decays.

Since the inflaton decays provide the leptons in terms of the number density divided by the entropy density as $\sim 3B T_R/4m_\f $, $\Delta_N$ to entropy density  is given as
\begin{align}
{\D_N\over s}&\simeq \frac{3}{4}\frac{T_R}{m_\f}B \times \(P_{L_\f\to L_N}-P_{\overline{L}_\f \to \overline{L}_N} \)\non \\ \non
&\simeq \frac{3}{4}\frac{T_R}{m_\f} B \xi_{CP} {|y_N|^2 \over 4\a_2^2}\\
&= 10^{-10} \frac{T_R}{m_\f} B \xi_{CP} \(\frac{|y_N|}{10^{-6}}\)^2. \laq{asy}
\end{align}
Here we have defined 
$$
\xi_{CP}\equiv \frac{c_\t  {y_{N_\t} \sum_{i=e,\m}{\Im[c^*_i] y_{N_i}}}}{|y_N|^2}.
$$
We stress here that this value of generated asymmetry does not depend on $T_R$ once $T_R/m_\f$ is fixed. 
This implies that the reheating temperature has no restriction in generating $\D_N$.  

Since $\Delta_N=-\Delta_\text{vis}$, the non-zero $\D_N$ means that there exists
\beq
\frac{\D_{\rm vis}}{s}\sim -10^{-10} \frac{T_R}{m_\f} B \xi_{CP} \(\frac{|y_N|}{10^{-6}}\)^2.
\eeq
This is transferred into the baryon asymmetry via the sphaleron process.
The required value of the lepton asymmetry converted from the measured
baryon asymmetry of the
universe~\cite{Ade:2015xua,Klinkhamer:1984di,Kuzmin:1985mm} is
\beq
\laq{reqas}
\(\frac{\D_{\rm vis}}{s}\)^{\rm required}=-(2.45\pm0.01)\times10^{-10}.
\eeq
Comparing with Eq.~\eqref{eq:asy}, we see that enough amount of baryon
asymmetry can be generated just after the reheating.  The question is
whether this asymmetry remains until today.

Let us consider the condition for preserving the baryon
asymmetry until today.  Obviously, $\D_N$ should not be transferred back
to the visible sector via Eq.~\eq{1} until the sphaleron process becomes
inefficient at the temperature lower than $T_{\rm sph}$.
Otherwise, the sphaleron would washout the baryon asymmetry.  Therefore,
$N$ should not be thermalized until $T=T_{\rm sph}.$ The interaction
rate of relativistic $N$ with the thermal plasma is given by
\beq
\G_N^{\rm th}\simeq \gamma_{N} |y_N|^2 T
\eeq
where $\gamma_{N}\simeq0.01$ is the numerical result from Refs.~\cite{Besak:2012qm, Hernandez:2016kel, Ghiglieri:2017gjz} which includes $2 \leftrightarrow 2$ and $1 \leftrightarrow 2$ processes as well as the LPM effect. 
By comparing  $\G_N^{\rm th}$ with the Hubble parameter at the radiation dominant era, $H\simeq \sqrt{g_* \pi^2 T^4/90M_P^2},$ one obtains the temperature that $N$ is thermalized 
\beq
\laq{thnt}
T\lesssim T_{\rm th}^N\simeq 7\TEV \(\frac{|y_N|}{10^{-6}}\)^2.
\eeq
The thermalization of $N$ can be avoided if we take into account its Majorana mass parameter, $M_N$, as
\beq
\d{\cal L} = -{M_N\o 2} \bar{N^c} N 
\eeq
which satisfies
\beq
M_{N}\gtrsim T^{N}_{\rm th}. 
\eeq
In this case, before the thermalization occurs $N$
becomes non-relativistic so that the asymmetry, $\D_N$, is washed-out
while the produced baryon asymmetry corresponding to $\D_{\rm vis}$
untouched.  On the other hand, $T_R\gtrsim M_N$ is necessary for our
discussion, i.e. the ``observation'' of active states with producing $N$
has to be valid kinematically. One arrives at the condition for
our scenario in inflaton perturbative decay
\beq
\laq{cond}
T_R\gtrsim M\gtrsim T_{\rm th}^N \simeq 7\TEV \({|y_N|\o 10^{-6}}\)^2. 
\eeq
This condition predicts specific patterns of the both active and right-handed neutrino masses and the relating phenomena, as we shall see soon.

\subsection{Implications on neutrino physics}
Since  $N$ is the right-handed neutrino, it gives a mass of active neutrino through the type-I 
seesaw mechanism,
\beq
\laq{ss1}
 \d m_\nu=\frac{| y_N|^2 \vev{H}^2}{M_N}. 
\eeq
Here $\vev{H}\simeq 174\GEV$ is the Higgs vacuum expectation value.
Substituting the condition for the baryogenesis \eq{cond} to Eq.~\eq{ss1} one can estimate the active neutrino mass as 
\beq
\laq{lnu}
  \d m_\nu \lesssim \frac{|y_N|^2 \vev{H}^2}{T_{\rm th}^N} \simeq 4 \times 10^{-3} \EV ~~~({\rm for}~M_N \gtrsim T_{\rm th}^N).
\eeq
Compared to the neutrino mass scales, $\sqrt{\Delta
m_\text{sol}^{2}}\simeq9\times10^{-3}~\text{eV}$ and $\sqrt{\Delta
m_\text{atm}^{2}}\simeq5\times10^{-2}~\text{eV}$, the $N$ particle which
is responsible for baryogenesis can significantly contribute to the
active neutrino masses only for the lightest or the second lightest
ones.
Two other right-handed neutrinos need to explain the rest of the
neutrino masses.

Based on the above discussion, the baryogenesis scenario predicts the
active neutrinos in either normal hierarchy (NH) or inverted hierarchy
(IH), i.e. not degenerated.
The sum of the active neutrino masses is determined for each mass
hierarchy, $\sum_I m_{\nu_I} \simeq 0.06~(0.10)~{\rm eV}$ for the NH
(IH)\footnote{For the estimation of the total neutrino mass we use
results in a global analysis of neutrino oscillation
measurements~\cite{Esteban:2018azc}.} with $I=1,2,3$ denoting the generation of active neutrinos in
the mass basis.  
The sum of masses has been constrained by the observations of cosmic
microwave background (CMB) and baryonic acoustic oscillation given as
$\sum_I m_{\nu_I}<0.12$ eV \cite{Aghanim:2018eyx}. 
The value is consistent with Eq~\eqref{eq:lnu}. The future observations
should improve the upper bound so that the scenario can be tested.

The prediction on the lightest neutrino mass in
Eq.~\eq{lnu} impacts on neutrinoless double beta decay.  Its decay rate
is characterized by the effective neutrino mass $m_\text{eff}$, whose
definition is $m_\text{eff}=\sum_{I} m_{\nu_I} [U_\text{PMNS}]_{e
I}^{2}$. 
Since the lightest neutrino mass is at most $|\delta m_{\nu}|$, we find
\beq
 |m_\text{eff}| \lesssim 7\times10^{-3}~{\rm eV}~~{\rm for~NH}~~{\rm and}~~0.01~{\rm eV} \lesssim |m_\text{eff}| \lesssim 0.05~{\rm eV}~~{\rm for~IH}.
\eeq

The masses of other two right-handed neutrinos are restricted since they should not wash out the lepton asymmetry $\D_{\rm vis}$.  
Here for simplicity we restrict ourselves in the case that the reheating temperature is so low, e.g. $T_R\lesssim 100\TEV$, that all the interaction rates via Standard Model Yukawa couplings are faster than the expansion rate.\footnote{If this assumption is removed, some of the charged lepton Yukawa coupling can be neglected when the washout is effective. There can be flavor-dependent lepton symmetry, 
and thus some component of $\D_{\rm vis}$ can not be washed out. 
In this case, there can be mass patterns where $M_2$ or $M_3$ is below $T_R$.  The extension is straightforward.} 
Under this most dangerous circumstance for the wash out,
one can obtain four possible mass patterns to evade the wash out:
\begin{align}
\laq{C1}
{\rm Case~1:}&~~~M_{2,3} \ll 100 \GEV, \\ 
\laq{C2}
{\rm Case~2:}&~~~T_R\ll M_{2,3}, \\
\laq{C3}
{\rm Case~3:}&~~~M_{2} \ll 100\GEV \ll T_R\ll M_3, \\
\laq{C4}
{\rm Case~4:}&~~~M_{3} \ll 100\GEV \ll T_R\ll M_2.
\end{align}
The Case 1 says that the masses of $N_{2,3}$ that violates the lepton number are almost negligible at $T>T_{\rm sph}.$
Case 2 is the possibility that $N_{2,3}$ are so heavy that the thermal production are kinematically suppressed, and that they are hardly thermalized. 
Case 3 or 4 is the composition of Cases 1 and 2. Apart from Case 2 there exists the right-handed neutrino with the mass below the electroweak scale.

Such a light particle may be probed experimentally.
The right-handed neutrinos in the mass range of $\mathcal{O}(1{\mathchar`-}10)~{\rm GeV}$ will be searched for in future beam-dump and collider experiments e.g. Ref.~\cite{Alekhin:2015byh}. 
Moreover such a particle can impact on the neutrinoless double beta decay as an additional intermediate state and thus may be tested indirectly~\cite{Benes:2005hn}.\footnote{If two degenerate right-handed neutrinos affect the process simultaneously the contribution to the decay rate is always destructive~\cite{Asaka:2011pb}.} The contribution behaves as $M_\a^{-2}$ because of a suppression of nuclear matrix element when the mass is larger than its typical momentum exchange $\sqrt{\vev{p^2}} \sim 200~{\rm MeV}$.

\subsection{Numerical estimation}
Here we perform a numerical simulation to confirm the discussion on the asymmetry separation. See Ref.\,\cite{Hamada:2018epb} for the detail analysis.
We focus on the two components of the density matrices:
\beq
\left(\r_{\bf k}\right)_{ij}=
\int_{|{\bf p}| \sim |{\bf k}|} {d^3 {\bf p}\over (2 \pi)^3}
\, 
{\r_{ij}({\bf p},t) \over s },
\ebq 
\left(\d \r_T\right)_{ij}=\int_{\ab{{\bf p}}\sim T}{{d^3 {\bf p}\over (2
\pi)^3}
\(
{\r_{ij}({\bf p}) \over s}
-
{\r^{\rm eq}_{ij}({\bf p}) \over s}
\) 
},
\eeq
and those for anti-leptons. Here $s$ is the entropy density. 
The first component, $\rho_{\bf k}$, represents the energetic leptons
produced by the $\f$ decay with initial typical momentum,
$
 |{\bf k}| =
m_\phi.
$
The second component, $\delta \rho_T$, represents leptons that deviate from the thermal distribution
with the typical momentum $|{\bf p}| \sim T$. Here
$T\simeq T_R$ is the temperature, and 
$\r_{ij}^{\rm eq}={ \d_{ij} / ( e^{|{\bf p}|/T}+1} )$ represents
the density matrix in the thermal equilibrium, which denotes the preexisting thermal plasma. 
We did not write down the equation for the right-handed neutrino since  $\D_N=-\D_{\rm vis}$ is guaranteed and we will estimate $\D_{\rm vis}.$

The time evolutions of the matrices can be obtained by solving the kinetic equations, which are derived from first principle with approximations~\cite{Sigl:1992fn}. The equations are given as
\begin{align}
  i\frac{d \rho_{ {\bf k}}}{dt} = [\Omega_{ {\bf k}} , \rho_{ {\bf k}}] - 
  \frac{i}{2} \{ \Gamma_{ {\bf k}}^d, \rho_{ {\bf k}}  \},
\label{eq:kinK}
\end{align}
\begin{align}
  i\frac{d \d \rho_T}{dt} = [\Omega_{ T} , \d \rho_{ T}] - 
  \frac{i}{2} \{ \Gamma_{ T}^d, \d \rho_{ T}  \}+{i} \d \G_{T}^p ,
\label{eq:kinT}
\end{align}
where $\Omega_{\bf k} =E_i(|k|) \d_{ij} $ and $\Omega_{ T} =E_i(T) \d_{ij} $.
The destruction and production rates for leptons are given by
\begin{align}
\left(\Gamma_{ {\bf k}}^d\right)_{ij}
	&\simeq
C \a_2^2  T \sqrt{\dfrac{T}{|{\bf k}|}} \d_{ij}
\label{eq:GammaK}
\end{align}
\begin{align}
\left(\Gamma_{ T }^d\right)_{ij} \simeq
C' \a_2^2T \d_{ij}
+\gamma_L T(\d_{i\tau}\d_{\tau j} y_\tau^2
 +\d_{i\mu}\d_{\mu j} y_\m^2 +\d_{ie}\d_{e j} y_e^2 )
+\gamma_N T y_{N_i}y_{N_j},
\label{eq:GammaT}
\end{align}
\begin{align}
\left(\d  \G_T^{p}\right)_{ij}
	&\simeq
C \a_2^2  T \sqrt{\frac{T}{|{\bf k}|}} \left(\r_{\bf k}\right)_{ij}
-C' \a_2^2T\left(\d \ol{\r}_T
\right)_{ij}.
\label{eq:deltaGamma}
\end{align}
The equations for the anti-leptons are obtained by replacing $ \rho$
with $\bar \rho$ everywhere and reversing the sign of $\Omega$'s.
In the actual numerical computation, the kinetic equations of
right-handed leptons and the red-shift of momenta are taken into account~\cite{Hamada:2018epb}.

\begin{figure}[!t]
\begin{center}  
   \includegraphics[width=105mm]{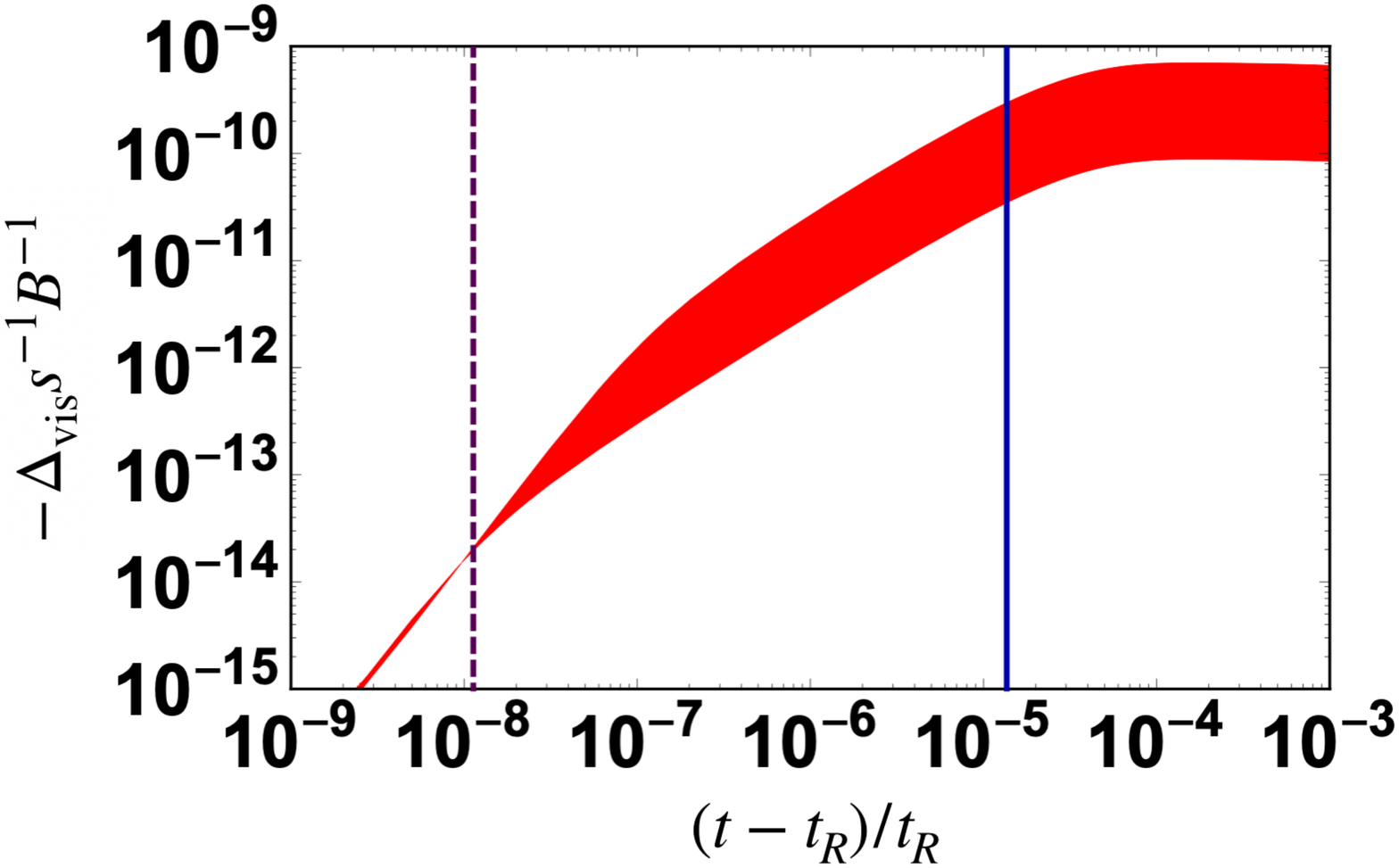}
      \end{center}
\caption{
The time evolution of the lepton asymmetry with
 $c_i={1\over\sqrt{3}}(\exp{i},\exp{2i},1)$, where $B=1, m_\phi=T_R= 10^7\GEV, y_N=10^{-6}(1,1,1)$ are taken. The red band corresponds to the variance of $C=C'$ between $1/3$ and $3$. The blue solid and purple dashed lines are the time scale of $t_{\rm MFP}$ and $1/(\g_L y_\tau^2 T_R),$ respectively, at which the pair annihilation and the scattering with the tau Yukawa coupling are important. }
\label{fig:1}
\end{figure}

Now let us briefly explain the terms in the rates.
The terms with the coefficient $C, C'$ corresponds to the $\G_{\rm th},\G_{\rm pair}$  respectively.  
We put a parameter $C, C' = {\cal O}(1)$ to take into account the theoretical uncertainty in the LPM effects and  in the energy distributions of the inflaton decay product. 
The terms with $\gamma_L, \gamma_N$ describe the scattering, and (inverse) decay via the Yukawa interactions of charged lepton and \eq{1}, respectively, which are important for ``observing'' the flavor. 
Here, we take $$\gamma_L = \gamma_N$$because of the same kinematics and gauge structure if we neglect $g_Y$.  
This term with $\gamma_N$ divides the lepton asymmetries into the two sectors. 
The numerical result of $\g_N$ can be found from~\cite{Besak:2012qm, Hernandez:2016kel, Ghiglieri:2017gjz}. We have neglected the scattering via Yukawa interactions for the high energy mode since it is much slower than the energy loss process.

We have approximated that the Higgs bosons are in thermal distributions, which is justified as follows. 
If the asymmetry of Higgs produced by the CP-violating effect is transferred into the right-handed neutrinos, the contribution is suppressed by $\O(y_N^4).$
On the other hand, most left-handed leptons produced from the interaction with Higgs bosons are in flavor eigenstates, and that the flavor oscillations are suppressed. 
Thus we can safely neglect the out-of-equilibrium effects of the Higgs bosons.

Now we are ready to  solve the kinetic equations.
The initial conditions of the density matrices for our scenario are as follows
\begin{align}
&\r_{\bf{k}}|_{t=t_R}=\ol{\r}_{\bf{k}}|_{t=t_R}= {{3\over 4}{  T_R \over m_\f} Bc_i^*  c_j },
\quad \d \r_T|_{t=t_R} = \d \bar \r_T|_{t=t_R}= 0.
\label{eq:initial}
\end{align}
We have assumed the absence of the deviation from the thermal equilibrium for the preexisting thermal plasma at $t=t_R.$
We take $\r_{\bf {k}}$ corresponding to \Eq{ph}.

In fig.~\ref{fig:1}, we show the asymmetry $\D_L/(sB)$ by varying the Hubble time $(t-t_R)/t_R$ with $T_R=10^7\GEV$, $m_\f=T_R,$ and $\g_N=\g_L=0.01.$ 
The vertical purple dashed and blue solid lines represent the time scale of $t_{\rm MFP}$ and that the density matrices 
are collapsed into the flavor eigenbasis due to the charged $\t$-Yukawa interaction, $(\g_L y_\t^2 T)^{-1}$. 
One finds that the asymmetry production lasts much shorter than $t_R\simeq 1/H|_{t=t_R}$, and the behavior around the timescales are consistent with the discussions in the previous section.
We have also checked that the amount of asymmetry does not change much by changing the reheating temperature.

\section{Case with reheating via dissipation processes}
\lac{Sec3}
Since the mechanism discussed previously is tied to the inflaton sector, for certain reheating dynamics, the asymmetry can be enhanced significantly. 
In what follows, we consider the reheating scenario where $ T_R /m_\f\gg \O(1)$.  For the inflaton perturbative decays, it was a thermal blocking effect that prevents the $T_R$ from going beyond $m_\f.$ This is because the decays of $\f$ to the daughter particles with thermal mass $\sim T_R\gtrsim m_\f$ are kinematically forbidden. 

When the thermal blocking effect is important, a thermal dissipation effect is also important~\cite{Yokoyama:2005dv,Anisimov:2008dz,Drewes:2010pf,Mukaida:2012qn,Drewes:2013iaa,Mukaida:2012bz}. 
If the dissipation effect is efficient, the reheating proceeds through the scatterings among the inflaton condensate and preexisting thermal plasma. 
The process can be represented as 
\beq
\laq{dis}
\f+Y \to X+L_\f.
\eeq
For instance, one may take $Y={\rm a~ Higgs~boson}$, $X=$ a right-handed charged anti-lepton.
This  process is kinematically allowed even if $m_\f$ is much smaller than the thermal masses of particles. This implies the parameter region of 
\beq T_R\gg m_\f\eeq  
is also possible.
In particular, the inflaton condensate loses energy of around $m_\f\ll T_R$ per one scattering. This means when the energy of the inflaton condensate all becomes the radiation, the scatterings take place $ 3T_R /4m_\f \gg 1$ times in a unit volume. 
Therefore, in this scenario $3T_R /4 m_\f $ leptons carrying momenta $$|{\bf p}|\sim T_R$$ are produced. They are out of equilibrium and thermalized after undergoing the flavor oscillation as in the previous part. 
As a result the same formula of \eq{asy} is expected, but with $T_R/m_\f\gg1.$
In particular, when $T_R/m_\f\gtrsim 10^2$  the baryon asymmetry can be explained with 
\beq
\laq{YNdis}
|y_N| \lesssim 10^{-7}.
\eeq
Thus,
\beq
\laq{cond2}
T_{\rm th}^{N}\lesssim T_{\rm sph}
\eeq
can be satisfied. This means that $\D_N$ is not transferred back to the visible sector until the sphaleron freezes out. The baryon asymmetry remains until today. Consequently,
our scenario works for the reheating temperature satisfying
\beq
T_R\gtrsim T_{\rm sph}\sim 100\GEV. 
\eeq
In this case, $N$ can be identified with any of the three right-handed neutrinos.

Let us discuss this possibility in more detail by introducing ALP inflation models~\cite{Daido:2017wwb, Daido:2017tbr, Takahashi:2019qmh} where the inflaton, $\f$, is an axion-like particle (ALP). The inflation scale can be as high as $\L_{\rm inf}\simeq 100\GEV-100\TEV$.\footnote{The QCD axion window can be opened and the moduli problem can be alleviated due to the low-scale inflation if inflation lasts long enough and if no mixing between the inflaton and the axion~\cite{Graham:2018jyp, Guth:2018hsa,Ho:2019ayl}. If there is a mixing which shifts the axion phase by $\pi$, the QCD axion can be set on the hilltop and thus a heavier QCD axion dark matter than usual is also possible~\cite{Takahashi:2019pqf}. }
The ALP (effective) mass\footnote{For the ALP miracle scenario~\cite{Daido:2017wwb, Daido:2017tbr} $m_\f$ should be identified as the effective mass of the inflaton. The inflaton mass at the vacuum, on the other hand, is highly suppressed due to an upside-down symmetry so that $\f$ is long-lived. } 
 $m_\f\sim \L_{\rm inf}^2/f$ and decay constant, $f$ have typical relation fixed by the CMB normalization of the primordial density perturbation:
\beq
m_\f \sim 10^{-6}f.
\eeq

The flavor oscillation occurs by introducing flavor-dependent couplings of $\f$ to leptons responsible for the reheating. 
The couplings are given as
\beq
{\delta \cal L}\simeq {i\f\over f} \sum_{ij}c_{ij} y_j H^* \bar{e}_i \hat{P}_L L_i,
\eeq
where $c_{ij}$ are dimensionless constants related with derivative $\f$-lepton couplings since $\f$ is an ALP. 
The reheating through the kind of couplings is shown to be successful in Refs.~\cite{Daido:2017tbr, Takahashi:2019qmh} if the $\f$ coupling to $\tau$ is large enough. 
This coupling also contributes to the dispersion relations and the scattering rates of the leptons, while the contributions, which are suppressed by $\O\((y_iT_R/f)^2\)$, are negligible. 
Thus \Eq{asy} holds with $\ab{{\bf p}}\sim T_R \sim 10^{2-3}m_\f$, $B=1$, 
and $c_i$ to be an eigenvector of $c_{ij}.$
The reheating occurs instantaneously, i.e. after inflation the energy density of the inflaton promptly becomes the radiation, $T_R \sim \L_{\rm inf},$ if $f\lesssim 10^8\GEV$.
As a result, the scenarios predict \beq T_R\sim 10^{2-3} m_\f .\laq{rate}\eeq 

The phenomenological implications are as follows.
The flavor mixing for the flavor oscillation leads to the process $\tau \to \mu/e +\f$  if kinematically allowed. This process can be searched for in Belle II experiment~\cite{Daido:2017tbr}. 
 $|y_N|\sim 10^{-8}-10^{-7}$ is predicted from  \Eqs{asy}, \eq{reqas} and \eq{rate}. 
The right-handed neutrino mass is 
\beq
M_N \simeq 6\GEV \({|y_N| \over 10^{-7}}\)^2\({0.05\EV \over \d m_\n}\).
\eeq
The beam-dump and collider tests, as well as the enhancement of the neutrinoless double beta decay rate, are interesting as discussed previously. 
The dark matter candidate may be the inflaton itself if the inflaton potential has an upside-down symmetry~\cite{Daido:2017wwb, Daido:2017tbr}. The dark matter can be searched for in the IAXO experiment~\cite{ Irastorza:2011gs, Armengaud:2014gea, Armengaud:2019uso}.  In this case the lightest right-handed neutrino may be the candidate as well.

\section{Discussion and Conclusions}

We have discussed a mechanism of baryogenesis through the CP violation
in the flavor oscillation in the reheating era.
Since a part of the lepton asymmetry is distributed to the right-handed
neutrinos through the Yukawa interactions, one obtains asymmetry in the
Standard Model sector, which is converted into the baryon asymmetry by the
sphalerons.
The scenario works for the reheating temperature of the Universe greater
than $\O(10)\TEV\OR \O(100)\GEV$ depending on the dynamics of inflaton.
The baryogenesis mechanism can be compatible with various new
physics models that favor low reheating temperature, such as
supersymmetric model avoiding gravitino problem, or require low
reheating temperature due to the low cutoff scale, e.g. relaxion models,
models with large extra-dimensions, composite Higgs models.

We  note
that the mother particle producing the left-handed leptons may not be
the inflaton, but the moduli, or heavy fermions that once dominate the
Universe. Even in those cases, the mechanism works. One can also apply the mechanism to the asymmetric
dark matter scenario~\cite{Barr:1990ca, Kaplan:1991ah, Kitano:2004sv}. 
By assuming that the matter couples to the dark matter with baryon
(lepton) number preserving interaction, the matter from the inflaton decay undergoes the
flavor oscillation caused by the misalignment of the oscillation and
interaction basis. The matter-antimatter asymmetry is distributed to the 
dark matter due to the CP-violating oscillation. 

We have discussed the dissipation effect which enhances the produced
lepton asymmetry. This effect can also enhance the asymmetry production of the scenario in Ref.~\cite{Hamada:2018epb}.  For
$T_R/m_\f\gtrsim\O(10^3),$ $T_R\lesssim 10^{7}\GEV$ is possible to explain
the baryon asymmetry of the Universe.

\section*{Acknowledgements}
WY thanks the KEK theory center for hospitality when this work was initiated. RK and WY also would like to thank the theory group at UC Davis for hospitality during their stay. 
This work is supported by JSPS KAKENHI Grant No.~15KK0176 (RK), 19H00689
(RK), and MEXT KAKENHI Grant No.~18H05542 (RK) and by NRF Strategic
Research Program NRF-2017R1E1A1A01072736 (WY).

\end{document}